# Observation of Dual-band Topological Corner Modes in Acoustic Kagome Lattice with Long-range Interactions

Chen Chen[1], Tianning Chen[1], Wei Ding[1], Jian Zhu[1, 2*]

[1]*School of Mechanical Engineering, Xi'an Jiaotong University, Xi'an, Shaanxi 710049, P. R. China*

[2]*State Key Laboratory for Strength and Vibration of Mechanical Structures, School of Aerospace Engineering, Xi'an Jiaotong University, Xi'an, Shaanxi 710049, P. R. China*

**Abstract**

The recent exotic topological corner modes (CMs) in photonic higher-order topological insulators with long-distance interactions have attracted numerous attentions and enriched the physics than their condensed-matter counterparts. While the next-nearest-neighbour (NNN) coupling appears between NNN lattice sites unselectively, and the NNN coupling and their associated dynamics remains elusive in acoustics due to the waveguide-resonator model, an analogy of tight-binding model (TBM), drastically hinders its NNN coupling. Here, in acoustics, we demonstrate selective NNN coupling-induced CMs in split-ring resonators-based kagome crystal and observe dual-band CMs. Three types of CMs are demonstrated in the first bulk gap which can be explained by TBM considering NNN coupling and one type of CMs is observed in the second. All of these findings are verified theoretically and experimentally which reveals rich physics in acoustics, opening a new way towards tunable or multi-band metamaterials design, and offering opportunities for intriguing acoustic manipulation and energy localization.

**Introduction**

The discovery of topological states, which support non-dissipative energy transport, greatly revolutionized people's perception on physics and materials science (1-3). Recently, the research on topological states has been extended to classical wave systems, including the demonstration of quantum Hall-like effect (4-7), quantum spin Hall-like effect (8,9) and valley Hall-like effect (10-13) in acoustics and photonics. Distinct from the conventional first-order topological insulators (FOTIs), higher-order topological insulators (HOTIs) support topological states with dimensionality more than one lower than that of the bulk. After the first prediction of HOTIs based on tight-binding model (TBM) (14,15), quadrupole topological insulators (16-19) and Wannier-type topological insulators (20-27) with a hierarchy of topological states have been demonstrated in realistic physical platforms, which greatly enrich the research on topological states.

---

[*] Corresponding author
Email address: jianzhuxj@xjtu.edu.cn

HOTIs study at the early stage are mainly explained by the TBM with nearest-neighbour (NN) coupling. More recently, a new type of topological corner states, type II corner modes (CMs), was experimentally observed in photonic kagome lattice due to the next-nearest-neighbour (NNN) coupling (28-30) which was first introduced into Haldane model (31) to reveal quantum anomalous Hall insulator (QAHI) states. Although the recent works (32-34) have debates on the protected characteristics of CMs in breathing kagome lattice, the robust protection of CMs by the symmetry of the lattice have potential applications in energy localization. However, on the one hand, the NNN coupling can occur between all NNN lattice sites unselectively, resulting in limited strategy to tune the NNN coupling between the lattice sites and control the positions of CMs. On the other hand, to the best of our knowledge, the NNN couplings-based CMs and their associated dynamics have not been explored so far in acoustic systems because the commonest waveguide-resonator model (WRM) (20,21), an analogy of TBM in acoustics, drastically hinders the coupling between NNN lattice sites.

Here, we firstly theoretically and experimentally achieved the control on the presence of NNN coupling in the acoustic system by introducing split-ring resonators (SRRs) (35-39) into kagome lattice and observed dual-band CMs. Specifically, the introduction of SRRs into acoustic kagome lattices can split the first three bands of the band structures, forming two neighouring bulk gaps. The band-edge frequencies at K of these two bulk gaps can be continuously evolved by rotating SRRs with Dirac points appearance. In the first bulk band, another new type of CMs, type III CMs which can also be regarded as general bounded corner states (30,40) along with type II CMs, are observed, which can be predicted by the TBM with stronger NNN coupling. While there only exists conventional type I CMs in the second bulk gap and the positions of these CMs cannot be predicted only by the calculated Wannier centers. Besides, two types of new CMs based on the selective NNN coupling are observed by rotating SRRs to another nontrivial topological phase which enable to tune the position of the CMs. This work shows that the introduction of SRRs into kagome lattices can serve as an excellent materials platform for the investigation of NNN coupling-based CMs and reveal richer physical phenomena in the acoustic system. And the appearance of NNN coupling in kagome lattice demonstrates more intriguing acoustic localization phenomena in a tunable way.

**Acoustic model with SRRs-based Kagome Lattices**

The designed two-dimensional kagome lattice sonic crystal (SC) with each lattice site occupied by SRR with inner radius $r_1 = 4.5$ mm, outer radius $r_2 = 6$ mm and width of the split $w = 3$ mm is shown in Fig. 1A. The coupling strength of sound modes between SRRs can be controlled by $\theta$ and $d$. Where $\theta$ represents the rotation angle of the SRRs, $d = \sqrt{3}a/6$ is the distance between the SRRs and the center of the unit cell, $a = 35$ mm is lattice constant. The calculated dispersion curve of the system is shown in Fig. 1B, when $\theta = 0°$, the first three bands of the band structures are successfully split and two bulk gaps appear (black lines), located in two frequency ranges, i.e., 3481 Hz - 4421 Hz and 4603 Hz - 5163 Hz, respectively. Interestingly, these two bulk gaps can be closed when $\theta = 44°$ (blue dotted line) and $\theta = 90°$ (red dot dash line) with two Dirac points appearance. Further, the first threee orders of eigenfrequencies at K continuously evolving by rotating the SRRs is shown in Fig. 1C, where the bulk polarizations of each band with different $\theta$ are marked by different colors (see Section S1 for calculation details).

To explain the appearance of Dirac points at these two bulk gaps, we introduce rigid rods, owning radius same with the outer radius of SRRs, to construct an unperturbed kagome lattice as shown in Fig. 1D. The $C_3$ symmetry of the system of the K-point constrains the energy of eigenmodes of the first three bands at K concentrating on three different symmetry centers Q-point, P-point and O-point, which can be regarded as Q modes, P modes and O modes (Inset in Fig.1D). For this unperturbed kagome lattice, P modes and O modes are degeneracy because the scattering effect of these rigid rods on O-point and P-point are identical. While the shorter distance of P-point and O-point to these rods than that of Q-point causes the P and O modes are located in the higher frequency. For the SC in Fig. 1A, Q, P and O modes still exist because the $C_3$ symmetry of the system remains unchanged throughout the evolution of rotation angle, which frequencies are affected by the coupling of these modes with resonant modes of the SRRs. Further, we extracted the eigenmodes in Fig. 1C at different rotational angles as shown in Fig. 1E. When $\theta = 0°$, P modes can interact with resonant modes of SRRs maximally while O modes can interact with resonant modes of SRRs minimally. Thus, in this case, the frequency of O modes is higher than that of P modes. With the increasing of $\theta$ from 0° to 180°, the frequency of P modes increases to the maximum while that of O modes decreases to the minimum. When $\theta = 90°$, the increasing P modes and the decreasing O modes encounter and one Dirac point appears and the hybridized modes (P-O modes) form. This Dirac point can be regarded as a deterministic one (41) which is robust against the sizes of the SRRs. For Q modes, when $\theta = 0°$, the minimum coupling strengths between resonant modes and Q modes occur, which results in the highest frequency of Q modes. With the increasing of $\theta$ from 0° to 180°, the frequency of Q modes decreases to the minimum when $\theta = 90°$ and increases back to the maximum when $\theta = 180°$. At this evolution, the Q modes encounter with P modes at $\theta = 44°$ and with O modes at $\theta = 136°$, with two Dirac points appearing. Note that these two Dirac points are accidental Dirac points (41,42), which are sensitive to the sizes of the SRRs and even disappear forever (see Section S2 for analysis). Although the type of Dirac points in the first bulk gap and the second bulk gap are different, the SC undergoes topological phase transitions during the forming of these Dirac points by rotating SRRs operations. For convenience, in the first bulk gap, we refer SC with $\theta = 0°$ as SC-A with the calculated bulk polarization **P** (-1/3, -1/3), SC with $\theta = 90°$ as SC-B with **P** (1/3,1/3), SC with $\theta = 180°$ as SC-C with **P** (0,0).

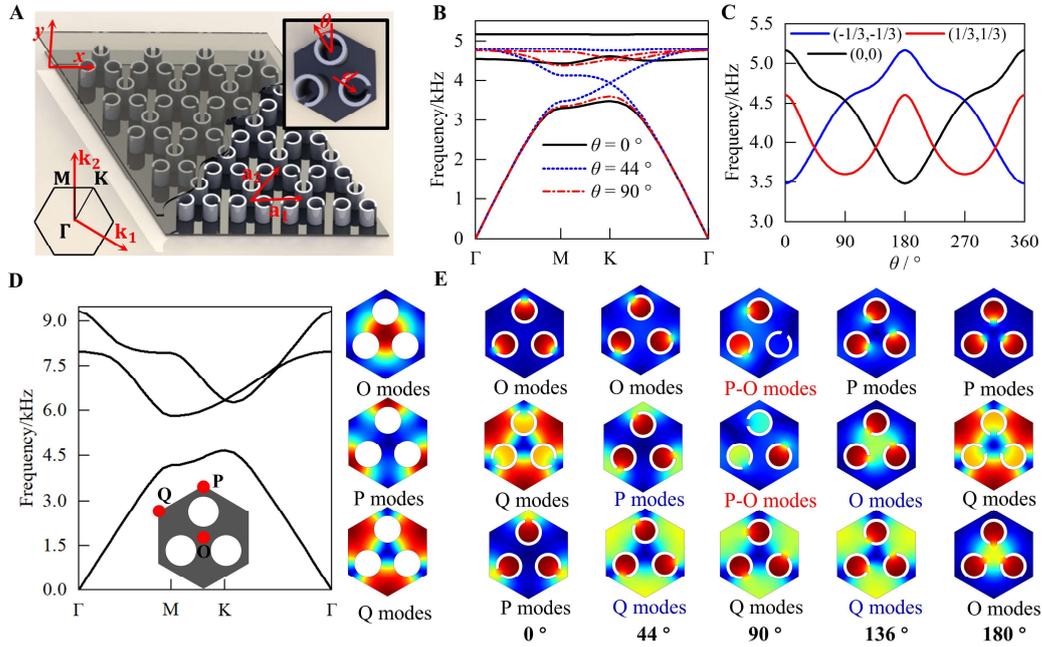

**Fig. 1. Acoustic model and topological phase transition.** (**A**) SC composed of acoustic kagome lattice with SRRs, inset: the corresponding unit cell (upper right) and the first Brillouin zone (lower left). (**B**) Band structures of the system when $\theta = 0°, 44°$ and $90°$. (**C**) The first three-order eigenfrequencies at K evolve with $\theta$ and the bulk polarization of each band is characterized by different colors. (**D**) Band structures and eigenmodes at K

for unperturbed kagome lattice consisting of rigid rods. (**E**) The evolution of eigenmodes at K with the rotation angle of the SRRs, where the deterministic degeneracy and accidental degeneracy are indicated by red and blue colors. Note that the frequencies of eigenmodes are increasing from lower panel to upper panel.

**CMs based on NNN coupling**

To investigate CMs of the nontrivial SC, an up-triangular super structure composed of SC-A (nontrivial SC) with the side length of $10a$ surrounded by SC-C (trivial SC) with the thickness of $3a$ is fabricated as shown in Fig. 2A, where the black triangle denotes the boundary between nontrivial and trivial SCs. Here, the outer trivial SC acts a role of open boundary condition in the TBM of the nontrivial SC, which is often used in photonic kagome lattice (28,29). Owing to the boundary cutting through the Wannier centers of the nontrivial SC, Type I CMs with energy localization in the three corner SRRs will appear in this super structure. As predicted in Fig. 2B, the calculated eigenfrequencies of this super structure at the first bulk gap demonstrate the triply degenerate type I CMs (red dots) at 3837 Hz appear in the gap of the bulk states. Interestingly, in addition to type I CMs, other two new CMs, type II CMs (blue dots) and type III CMs (magenta dots), can also be observed in the first bulk gap which appears in pairs, located in the two sides of the edge states due to the NNN coupling, and they own symmetric and antisymmetric field profiles about the bisectrix of the corner. These three CMs and their sound energy localization in the calculated field profiles are shown in Figs. 2C to 2E. The positions of sound energy localization are gradually away from the SRRs in the corner, which originates from the interactions between the edge states induced by NNN coupling (28, 30). Note that the type III CMs have not been observed before, because they need a stronger NNN coupling strength, which can be predicted by the TBM considering NNN coupling (see Section S3 for TBM analysis).

We experimentally verify the existence of these types of CMs by fabricating a sample as shown in Fig. 2A. The measured normalized sound transmission is demonstrated in Fig. 2F, where the peak of each transmission spectrum presents the measured eigenfrequency of the corresponding CMs. The measured sound pressure amplitudes for the cavities of the SRRs near the corner for type I CMs, anti-symmetric type II CMs and anti-symmetric type III CMs are shown in the insets in Figs. 2C to 2E, where the stars indicate the positions of the sources and the red and black stars are sources with reverse phases to excite anti-symmetric modes. However, the absolute pressure profiles cannot present the symmetry of the CMs along the bisectrix of the corner. We therefore extracted the time-domain signals at the cavities of the corresponding SRRs to show the symmetry of the CMs along the bisectrix of the corner (see Section S4 for analysis).

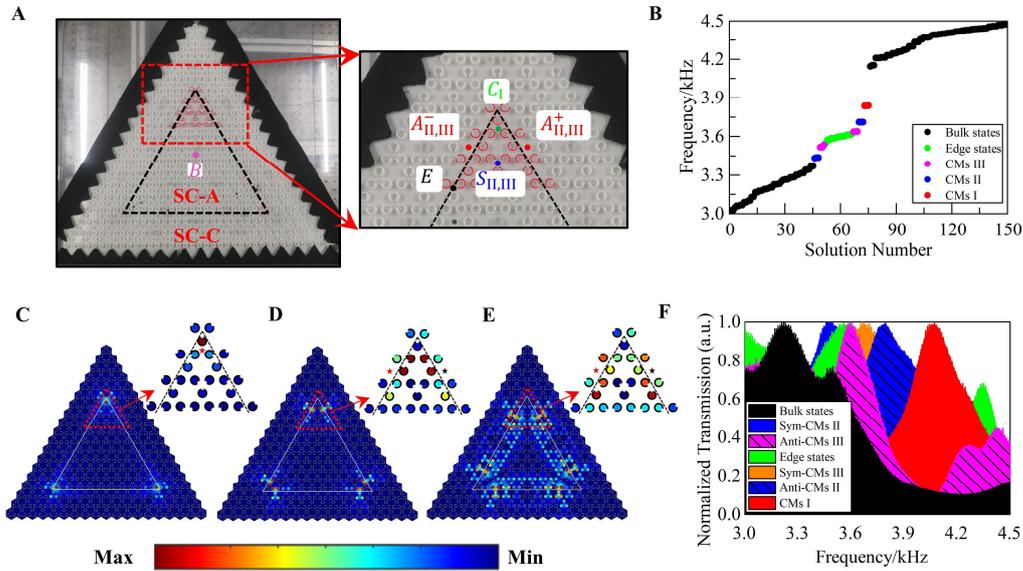

**Fig. 2. Simulated and experimental demonstrations of CMs.** (**A**) Photograph of the super structure composed of nontrivial SC-A with side length of $10a$ surrounded by SC-C (trivial SC) with thickness of $3a$, where the positions of the source are indicated by B (bulk states), E (edge states), $C_I$ (corner states I), $S_{II,III}$ (symmetric CMs II and III) and $A^{\pm}_{II,III}$ (two sources with reverse phase to excite antisymmetric CMs II and CMs III). (**B**) Simulated

eigenfrequencies distribution at the first bulk gap of the super structure. Simulated field profiles of type I CMs at 3837 Hz (**C**), antisymmetric type II CMs at 3710 Hz (**D**) and antisymmetric type III CMs at 3535 Hz (**E**). Insets presents the experimentally extracted pressure amplitudes at the cavities of the SRRs near the upper corner respectively and the stars denote the position of the sources. (**F**) Experimental measurements of normalized transmissions for different corner states, edge states and bulk states excitations.

## CMs in the second bulk gap

Interestingly, the CMs are observed not only in the first bulk gap but also in the second bulk gap. Here, we calculate the eigenfrequencies distributions of the super structure in the second bulk gap as shown in Fig. 3A, where only type I CMs appear in the bulk gap. The sound pressure profiles of the CMs are shown in Fig. 3B which indicates SC-A in the second bulk gap own nontrivial bulk polarization **P** (-1/3, -1/3). However, the calculated bulk polarization is equal to (0, 0), which is evaluated by the summation of bulk polarization of the first two bands. This is because the band with **P** (1/3,1/3) does not take part in the forming and breaking of the two Dirac points in the second bulk gap as shown in Fig. 1C. We do not need to consider the effect of the band when calculating the bulk polarization of the second band. Thus, actually, SC-A owns nontrivial bulk polarization **P** (-1/3, -1/3) in the second bulk gap. On the other hand, the eigenfrequencies of the corner states are close to the edge states, which result in the partial mixture of the corner states and edge states. To further split them, we can translate the SRRs to tune the couplings between SRRs (see Section S5 for details). The experimental measurements of normalized transmission in Fig. 3C agree well with the calculated eigenfrequency distributions in Fig. 3A, which indicates the appearance of type I CMs at 4962 Hz as shown in the red region. To further verify the appearance of CMs, the absolute pressure amplitudes of the cavities of the SRRs near the upper corner are extracted. As shown in Fig. 3D, the experimental measurement of the pressure profiles indicate that sound energy is well localized in the cavity of the SRR at the corner.

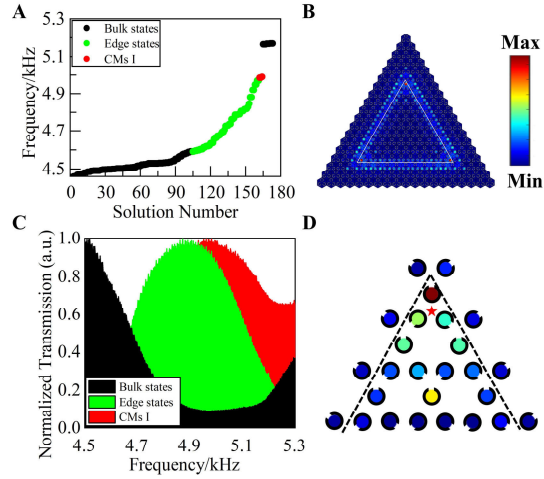

**Fig. 3. CMs in the second bulk gap.** (**A**) Eigenfrequencies distribution in the second bulk gap of the SC-A. (**B**) Simulated pressure profiles of the CMs at 4990 Hz. (**C**) Experimental measurements of normalized transmission for bulk excitation, edge excitation and corner excitation. (**D**) Experimentally extracted pressure amplitudes at the cavities of the SRRs near the upper corner for CMs.

## CMs induced by selective NNN coupling

Next, we investigate the CMs in SC-B ($\theta = 90°$) with nontrivial bulk polarization **P** (1/3, 1/3) by constructing an up-triangular super structure composed of SC-B with the side length of 10$a$ surrounded by SC-C (trivial SC) with the thickness of 3$a$ as shown in Fig. 4A. The simulated eigenspectra of this up-triangular super structure in Fig. 4B show the appearance of new type II CMs (blue dots) and type III CMs (magenta dots) in the bulk gap. The absolute pressure field of type II and type III CMs are demonstrated in Figs. 4C and 4D. Note that the upper-frequency type II CMs are antisymmetric modes while the type III CMs and lower-frequency type II CMs are symmetric modes. Here, the symmetric and antisymmetric modes mean the two lattice sites with the largest energy localization near the corner own the same and reverse phases, respectively.

To verify these types of CMs, we experimentally measure the normalized transmission at bulk excitation, edge excitation and corner excitations as shown in Fig. 4E, which is consistent with the simulated eigenfrequency distribution shown in Fig. 4B. The extracted absolute sound pressure amplitudes at the cavities of the SRRs near the upper corner for symmetric type II CMs and type III CMs are shown in the insets of Figs. 4C and 4D, where the red stars present the position of the source. Note that the measured positions of largest energy localizations in the inset of Fig. 4C are different from the simulated eigenmodes, an additional energy localization appears in position 2 due to the existence of environmental loss and is affected by the position of the source. The source can also excite the resonant modes of the SRRs oriented to it and the sound pressure amplitudes of these SRRs may be comparative big than those excited by the CMs due to the existence of environmental loss. The simulated pressure profiles considering loss can be found in Fig. S8 in Section S6, which demonstrates the same pressure profiles as experimental results.

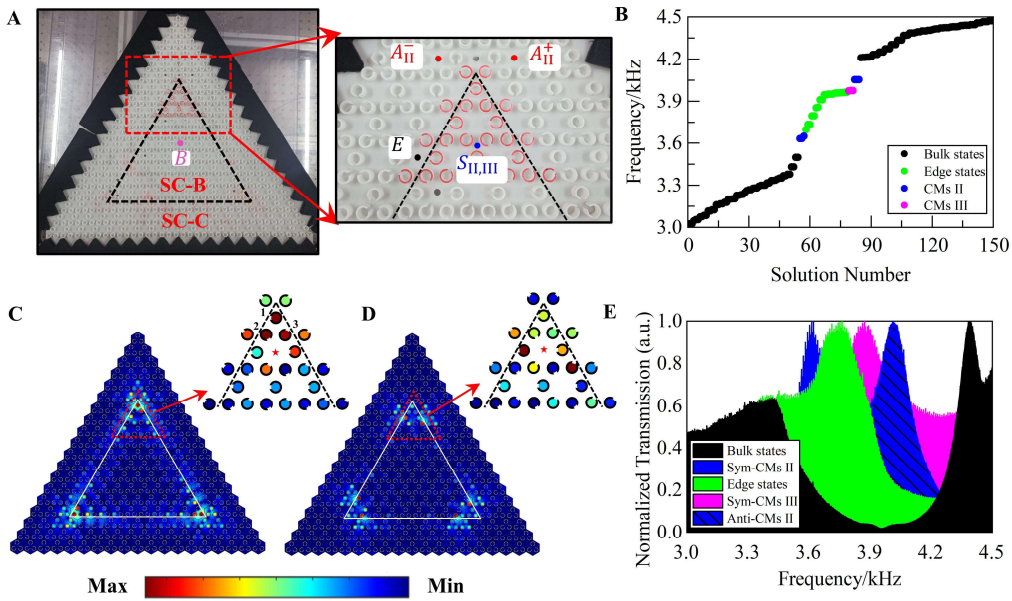

**Fig. 4. Simulated and experimental demonstration of CMs in SC-B.** (A) Photograph of the super structure composed of nontrivial SC-B with side length of 10$a$ surrounded by SC-

C (trivial SC) with thickness of $3a$, where the positions of the source are indicated by B (bulk states), E (edge states), $S_{II,III}$(symmetric CMs II and III) and $A^{\pm}_{II,III}$ (two sources with reverse phase to excite antisymmetric CMs II and CMs III). (**B**) Simulated eigenfrequencies distribution of the super structure. Simulated pressure field profiles of symmetric CMs II at 3649 Hz (**C**) and symmetric CMs III at 3975 Hz (**D**). Insets presents the experimentally extracted pressure amplitudes at the cavities of the SRRs near the upper corner respectively and the stars denote the position of the sources. (**E**) Experimental measurements of normalized transmission for bulk excitation, edge excitation and corner excitation.

We found that these all CMs demonstrated in SC-B can also be accurately predicted by TBM considering selective NNN coupling as demonstrated in Fig. 5A, which seems to be unique to this anisotropic SRRs system. Different from the system with $\theta = 0°$ where the NNN coupling exists between all NNN lattice sites unselectively, here the NNN coupling only appears between designate NNN lattice sites where their splits are oriented to each other. Specifically, the intra-cell hopping K and inter-cell hopping J are equal to each other because when $\theta = 90°$, the resonant and scattering effect of SRRs on P-point and O-point are equal. While the NNN hopping $J_1$ is selective between the SRRs (lattice sites) oriented to each other and bigger than K and J. When each side of the TBM contain 20-unit cells, the energy spectra of different eigenmodes evolving with the ratio of $J_1/K$ are demonstrated in Fig. 5B, where with the increasing of the $J_1/K$, symmetric (solid line) and antisymmetric (dashed line) type II CMs (blue line), type III CMs (magenta line) appear. Note that for SC-B, the antisymmetric type III CMs have mixed with the edge states, thus we cannot find it in the simulated and experimental results in Fig. 4. We further extracted the absolute pressure fields for symmetric type II CMs (Fig. 5C) and type III CMs (Fig. 5D) when $J_1/K=5$ and J=K, which are consistent with the simulated results in Fig. 4.

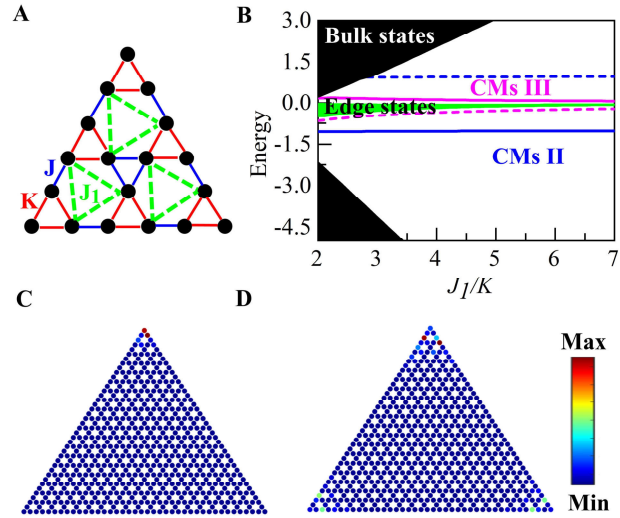

Fig. 5. CMs predicted by TBM in kagome lattice considering selective NNN coupling. (A) Schematic illustration of the TBM in kagome lattice considering selective NNN coupling strength $J_1$ when J=K. (B) The evolution of energy spectrum with ratio $J_1/K$ when the TBM contains 630 sites where the solid lines present the symmetric modes and dashed lines present the antisymmetric modes. The extracted absolute pressure field distributions of symmetric Type II CMs (C), symmetric Type III CMs (D) with $J_1/K$=5, J=K.

## Conclusion

The interactions between NNN lattice sites can lead to new CMs split from topological edges states, which was regarded to be specific to the electromagnetic system. Here, we observe this novel CMs in acoustic system and further prove the appearance of another exotic CMs with stronger NNN coupling. Besides, we demonstrate two types of CMs based on selective NNN coupling, which appear between designate NNN lattice sites not all NNN lattice site and are bigger than NN coupling. These acoustic CMs are observed in the same system but with different positions of energy localization. Specifically, we successfully split the first three bands of acoustic kagome lattice by introducing SRRs, which forms two neighboring bulk gaps. Interestingly, these two bulk gaps can be continuously opened and closed by rotating the SRRs with Dirac points' forming and breaking in the both first and second bulk gaps. The formation mechanism of these Dirac points is discussed in detail and these Dirac points divide the acoustic kagome lattice with three different topological phases in the first bulk gap and two different topological phases in the second bulk gap. For SC-A with nontrivial bulk polarization **P** (-1/3, -1/3) in the first and second bulk gap, three different types of CMs have been theoretically and experimentally demonstrated by considering NNN coupling in the first bulk gap. In addition to the type I and type II CMs, we discover a new type of corner states, i.e., type III CMs which require stronger NNN coupling strength in this nontrivial acoustic system. In the second bulk gap, we experimentally and theoretically demonstrate type I CMs. For SC-B with nontrivial bulk polarization **P** (1/3, 1/3) in the first bulk gap, CMs can also be predicted by TBM considering selective NNN coupling. The SC-C with trivial bulk polarization **P** (0,0) works as a cladding to surround the nontrivial system. Our work provides an efficient strategy to realize dual-band CMs and demonstrates richer energy localization phenomena by considering NNN coupling in acoustic systems. These novel phenomena open up a route to realize energy localization at different positions on-demand and provide the potential for.

**Acknowledgments**

This work was financially supported by the National Natural Science Foundation of China (Nos. 12002258 and 516750402), the Natural Science Foundation of Shaanxi Province (No. 2020JQ-043), the China Postdoctoral Science Foundation (No. 2018M643623) .